\pdfoutput=1
  
\documentclass[preprint,12pt]{elsarticle}
  
\usepackage{graphicx}

\usepackage{amssymb}
\usepackage{comment}

\usepackage[nodots,nocompress]{numcompress}

\journal{JQSRT}

\begin{document}

\begin{frontmatter}



\title{Cascade emission in electron beam ion trap plasma}


\author[label1]{Valdas Jonauskas}

\author[label1]{\v{S}ar\={u}nas Masys}
\author[label1]{Au\v{s}ra Kynien\.{e}}
\author[label1]{Gediminas Gaigalas}

\address[label1]{Institute of Theoretical Physics and Astronomy of Vilnius
University, 
\ A. Go\v{s}tauto 12, LT-01108  Vilnius, Lithuania}

\begin{abstract}

We present investigation of the influence of cascade emission to the formation of spectra from plasma created by electron beam ion trap (EBIT) in electron trapping mode. 
It has been shown that cascade emission can play an important role in the formation of spectra from the EBIT plasma. Process of the cascade emission takes place when ion having cycloidal orbit leaves electron beam where coronal approximation is applicable. Thus both processes -   excitation from ground or metastable levels and cascade emission - take part in the population of levels. Demonstration is based on the investigation of  $W^{13+}$ spectra.
The present investigation helps to resolve long-standing discrepancies; in particular, the present structure of $W^{13+}$ spectra is in good agreement with measurements on electron beam ion trap.
Lines in the experimental spectra are identified as $4f^{13} 5s 5p \rightarrow 4f^{13} 5s^{2}$ and $4f^{12} 5s 5p^{2} \rightarrow 4f^{12} 5s^{2} 5p$ transitions from Dirac-Fock-Slater calculations.

\end{abstract}


\end{frontmatter}


\section{Introduction}
An electron beam ion trap (EBIT) \cite{1988prl_60_1715_marrs} is an ideal tool to study radiation of highly charged ions as well as the interactions between electrons and ions - electron impact ionization, electron-ion recombination processes. Electron beam interacting with ions in an ion trap region produces atoms of element in the desirable ionization state which is determined by the energy of  electrons. Neighboring ionization states are less populated because the energy of electrons is too small to reach next ionization state or too large to stay in the current one. Consequently, the analysis for lines of spectra  can be focused on the main ionization states  obtained in the plasma. In addition, low density is another specific characteristic of the plasma. It is assumed that steady state is reached and population of levels does not change in time. Mainly collisional-radiative modeling is applied describing the population and depopulation of the levels. Collisional-radiative  modeling in the steady state 
conditions has been successfully used for the analysis of spectra from highly charged tungsten ions being trapped in the EBIT \cite{2006pra_74_042514_Ralchenko, 2007jpb_40_3861_ralchenko, 2007apmidf_13_45_radtke, 2008jpb_41_021003_ralchenko, 2008mnras_386_l62_chen}. However, it was shown that the population of the levels can be determined using coronal approximation \cite{Jonauskas2007jpb_40_2179}, i.e. only collisional excitation from the ground state needs to be considered. Good agreement has been obtained for line intensities when compared with data from collisional-radiative modeling.  On the other hand, cascade emission, as a general rule, is omitted in the analysis of spectral line formation for the EBIT plasma. We should note that term cascade emission is used here instead of radiative cascade in order to distinguish population of levels from higher levels through radiative cascade which is included in collisional-radiative model. As will be shown later, the population of levels differs for those two 
cases. The cascade processes have been mostly investigated for radiative and Auger decays of inner shell vacancies \cite{2003jpb_36_4403_jonauskas, 2008jpb_41_215005_jonauskas, 2010pra_82_043419_palaudoux, 2011pra_84_053415_jonauskas}.  However, trapped ions interact with electrons in the beam, but due to cycloidal orbits the ions can spend part of their time outside the electron beam \cite{currel2001, 1995ps_59_392_gillaspy}.   Process of the cascade emission has to start when interaction with electrons ends. This effect is more important for ions in the lower or medium  effective charge states \cite{2009apj_702_838_liang}. It was found that under same conditions, higher charge ions show less expansion in the radial direction. 
When the  charge state of ions increases the Coulomb's attraction force directed toward electron beam also increases which lead to the decrease of the time ions spend outside the beam where the cascade emission depopulates excited levels.
{ The role of cascade emission depends on the time the ions spend inside and outside the electron beam. However, this time is determined by many parameters: ion temperature, electron beam energy, electron beam current, electric and magnetic fields.  On the other hand, the time fraction ions moves in the electron beam can be approximately related to the electron beam diameter and radial distribution of ions in the trap. Effective electron density is often introduced in order to reduce ion-electron collision rates because ions do not dwell all the time in the electron beam region \cite{2009apj_702_838_liang, 2004apj_611_598_chen}. It was found that range of ion radius $r_{i}$ ratio against the geometric electron radius $r_{e}$ can be expressed through the effective charge $Z_{eff}$ of ion: $1/(Z_{eff}/Z)^{\alpha}$ with $1 < \alpha \leqslant 2$ \cite{2009apj_702_838_liang}. For $W^{13+}$, we have $r_{i}/r_{e} \approx 6^{\alpha}$, i. e. ions mainly reside outside the electron beam.}  

In this work we show the importance of cascade emission to the formation of spectra from EBIT plasma in electron trapping mode. To our knowledge, up to know the cascade emission in electron trapping mode has not been analyzed. Cascade emission has been investigated previously in magnetic trapping mode \cite{1996rsi_67_11_Beiersdorfer} observing X-ray emission after charge exchange of neutrals with highly charged ions. Magnetic trapping mode takes place after the electron beam is switched off, thus, electron-impact excitation does not influence the population of levels. 

Our demonstration for the role of cascade emission in electron trapping mode is based on the investigation of $W^{13+}$ spectra obtained at the Berlin EBIT \cite{Hutton_2003nimb_205_114, 2008_Wu_cjp_86_125}. 
Using the Berlin EBIT facilities Hutton et al. \cite{Hutton_2003nimb_205_114, 2008_Wu_cjp_86_125} measured spectra of the tungsten ions at electron beam energies corresponding to the maximum production of $W^{13+}$ ions in the EBIT plasma. They looked for resonance lines corresponding to the $5p \rightarrow 5s$ transition which could be candidates for fusion plasma diagnostics. It was expected that if collapse of $4f$ orbitals had occurred in the ion then two prominent resonance lines would be produced in the spectrum. In their work, two intense lines at 258.2 \AA{} and 365.3 \AA{} have been assigned to the $5p \rightarrow 5s$ transition. Their decisions were based on the agreement with their theoretical calculations, high intensities of lines in the spectra and dependence of line intensities on the electron beam energies. Theoretical analysis has been performed in single-configuration approximation using Cowan code \cite{Cowan_1981tass.book.....C} where quasirelativistic Hartree-Fock technique based on 
Schrodinger equation is adopted taking into account the most important one-electron corrections (mass-velocity and Darwin). 
Later the detailed investigation of transitions for $W^{13+}$ ion employing relativistic multireference many-body perturbation theory  has been performed in \cite{Trabert_2008jpcs_163_012017}. 
The calculations presented therein showed that ground state of the ion corresponds to the $4f^{13} 5s^{2}$ configuration and it does not coincide with earlier predictions obtained from quasirelativistic calculations \cite{Curtis_1980prl_45_2099} that Pm-like ions with a nuclear charge $Z$ larger than 73 have alkali-metal structure with a ground configuration of $4f^{14} 5s$.  Pm-like $W$ spectra simulated using collisional-radiative model for low density conditions did not help to identify the observed lines in the experimental spectra \cite{Trabert_2008jpcs_163_012017}. 
Vilkas et al \cite{Vilkas_2008pra_77_042510} stressed that their study does not show one or two dominant resonance lines in the spectra.  Thus, the situation with the interpretation of the lines from the EBIT plasma spectra corresponding to the radiative transitions in $W^{13+}$ ion was named as class of  {\it unsolved mysteries} \cite{2008_Wu_cjp_86_125}.

\section{Calculation of spectra}

Intensity of spectral lines is determined by equation:
\begin{equation}
I_{ij} = n_{i}  A^{r}_{ij} 
\label{e0}
\end{equation}
where $n_{i}$ is the population of level $i$; $A^{r}_{ij}$ - probability of radiative transition from level $i$ to level $j$.

Electron impact excitation takes place in the electron beam while the cascade emission occurs outside it. 
The initial population of the excited levels when ions leave the beam region can be found analyzing  excitation from the ground and metastable levels within coronal approximation framework.
For this, energy levels, radiative transition probabilities and excitation rates have to be calculated.

\subsection{Atomic data}

\begin{figure}
 \includegraphics[scale=0.5]{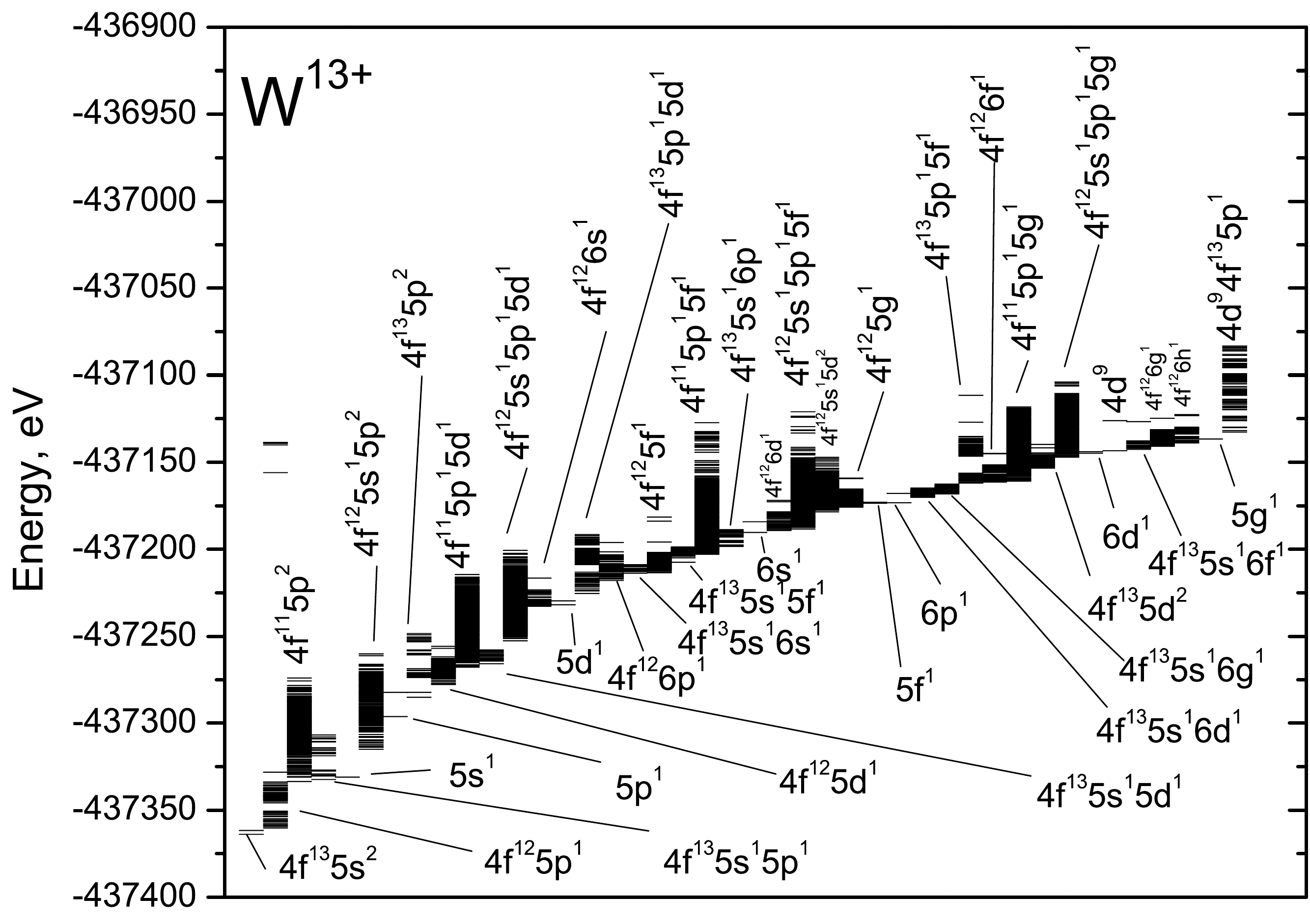}%
 \caption{\label{f0} Energy levels of $W^{13+}$ configurations.  }
 \end{figure}

Flexible Atomic Code 
\cite{Gu_2003ApJ...582.1241G} was employed to obtain energy levels and radiative transition probabilities in $W^{13+}$ using configuration interaction (CI) method. CI basis consists of 32661 levels which arise from 65 configurations: $4f^{13}   5s^{2}$, $4f^{14}  5s$, $4f^{14}  5l$, $4f^{14}  6l'$, $4f^{13}  5s 5l$, $4f^{13}  5s 6l'$, $4f^{12}  5s^{2}  5l$, $4f^{12}  5s^{2}  6l'$, $4d^{9}  4f^{14}  5s^{2}$, $4d^{9}  4f^{14}  5s 5l$, $4d^{9}  4f^{13}  5s^{2}  5l$, $4f^{12}  5s 5p^{2}$, $4f^{12}  5s 5p 5l''$, $4f^{12}  5s 5d^{2}$, $4f^{12}  5s 5d 5f$, $4f^{12}  5s 5d 5g$, $4f^{12}  5s 5f^{2}$, $4f^{12}  5s 5f 5g$, $4f^{12}  5s 5g^{2}$, $4f^{11}  5s^{2}  5p^{2}$, $4f^{11}$  $5s^{2}  5p 5l''$, $4f^{13}  5p^{2}$, $4f^{13}  5p 5l''$, $4f^{13}  5d^{2}$, $4f^{13}  5d 5f$, $4f^{13}  5d 5g$, $4f^{13}  5f^{2}$, $4f^{13}  5f 5g$, $4f^{13}  5g^{2}$ ($l=p,d,f,g$; $l'=s,p,d,f,g,h$; $l''=d,f,g$). 
Energy levels of the lowest configurations are shown in Fig. \ref{f0}.   Electric dipole transitions were calculated among all levels. However, analysis of magnetic dipole, quadrupole and electric quadrupole as well as octupole radiative transitions was performed for the five lowest configurations: $4f^{13} 5s^{2}$, $4f^{12} 5s^{2} 5p$, $4f^{11} 5s^{2} 5p^{2}$, $4f^{13} 5s 5p$ and $4f^{14} 5s$. Electron impact excitation rates to all levels were calculated from the ground level and the lowest levels of $4f^{12} 5s^{2} 5p$ configuration within distorted wave approximation. { Gaussian distribution function with a full width at a half-maximum of 30 eV used for electron energy.} Here we do not take into account ionization and recombination processes from and/or to other tungsten ions. Only levels belonging to the same ionization stage are considered. 

\subsection{Population of levels}

Population of levels after excitation from the ground level is calculated according to the formula:
\begin{equation}
n_{i} = \frac{N_{e} C_{0i}}{\sum_{k<i} A^{r}_{ik}},
\label{e1}
\end{equation}
where $0$ is the ground level index and  $i > 0$;  $C_{ik}$ - electron impact excitation/deexcitation rate from level $i$ to level $k$. 

Obtaining population of levels the additional term:
\begin{equation}
\frac{\sum_{k>i} n_{k} A^{r}_{ki}}{\sum_{m<i} A^{r}_{im}},
\label{e2}
\end{equation}
was added  in order to take into account supplement population of levels due to radiative cascade from higher levels.  Strictly speaking this term  is normally omitted in coronal approximation.

However, other population mechanisms appear on the scene when the ions leave electron beam region. The population of levels during the cascade emission changes in every step of cascade:
\begin{equation}
n_{i}^{j+1}=\sum_{l>i} \frac{ n_{l}^{j} A^{r}_{li}}{\sum_{k<l} A^{r}_{lk}},
\label{e3}
\end{equation}
where $n_{l}^{j}$ corresponds to the population of level $l$ in  $j$ step of the cascade. We  only consider  levels which have energies lower than ionization threshold and thus can not decay through autoionization transitions. 

Using branching ratio:
\begin{equation}
B_{il} = \frac{\sum_{k<i,k>l} A^{r}_{ik} B_{kl}}{\sum_{m<i} A^{r}_{im}}
\label{e4}
\end{equation}
the Eq. (\ref{e3}) can be written in such a way:
\begin{equation}
n_{i}=\sum_{l>i} n_{l}^{0} B_{li},
\label{e5}
\end{equation}
where $n_{l}^{0}$ is initial population of level $l$ before the beginning of the cascade emission. Since cascade emission takes place after ions leave electron beam, the initial population of the levels is found from Eq. (\ref{e1}) with Eq. (\ref{e2}) populating levels by electron impact from the ground or metastable levels.

In order to demonstrate the difference in level population for those two cases (Eq. (\ref{e1}) with Eq. (\ref{e2}) and Eq. (\ref{e3})), let's take a three level system. { The population $n_{1}$ of the first excited level obtained from Eq. (\ref{e1}) with Eq. (\ref{e2}) 
\begin{equation}
n_{1} = \frac{N_{e} C_{01}} {A^{r}_{10}} + \frac{n_{2}^{0} A^{r}_{21}} {A^{r}_{10}} = \frac{N_{e} C_{01}} {A^{r}_{10}} + \frac{N_{e} C_{02} A^{r}_{21}} {A^{r}_{10} (A^{r}_{21}+A^{r}_{20})}.
\label{e7}
\end{equation}
will be increased by the term due to cascade emission obtained from Eq. (\ref{e3})
\begin{equation}
n_{2}^{0} \frac{A^{r}_{21}}{A^{r}_{21}+A^{r}_{20}}=\frac{N_{e} C_{02} A^{r}_{21}} {(A^{r}_{21}+A^{r}_{20})^2}
\label{e6}
\end{equation}
where the initial population $n_{2}^{0}$ of the second excited level before emission cascade starts is expressed by
\begin{equation}
n_{2}^{0} = \frac{N_{e} C_{02}}{{A^{r}_{21}+A^{r}_{20}}}. 
\end{equation}
Thus, the cascade emission (Eq. (\ref{e6})) can very strongly increase the population of lower levels from higher populated levels, i. e. if $A^{r}_{10}$ is much larger than $A^{r}_{21}+A^{r}_{20}$.}

\section{Results and discussion}

\begin{figure}
 \includegraphics[scale=0.37]{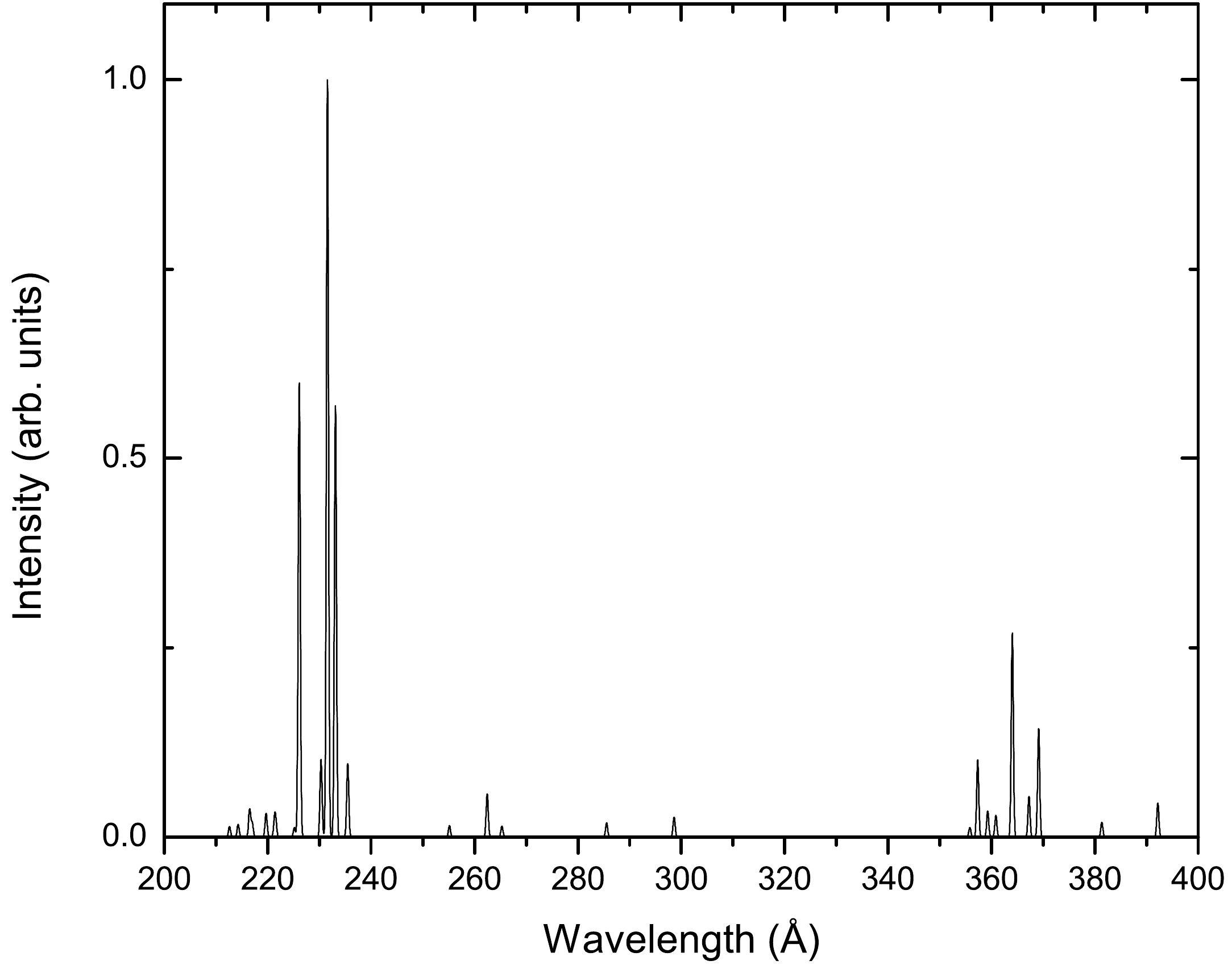}%
 \caption{\label{f01} { Theoretical spectrum obtained in coronal approximation at electron beam energy $E_{e}=300$ eV and electron density  $n_e=10^{12}$ cm$^{-3}$. Population of levels correspond to excitation from the ground level of $W^{13+}$ {(Eq.(\ref{e1}) with Eq. (\ref{e2}))}}. }
 \end{figure}

\begin{figure}
 \includegraphics[scale=0.3]{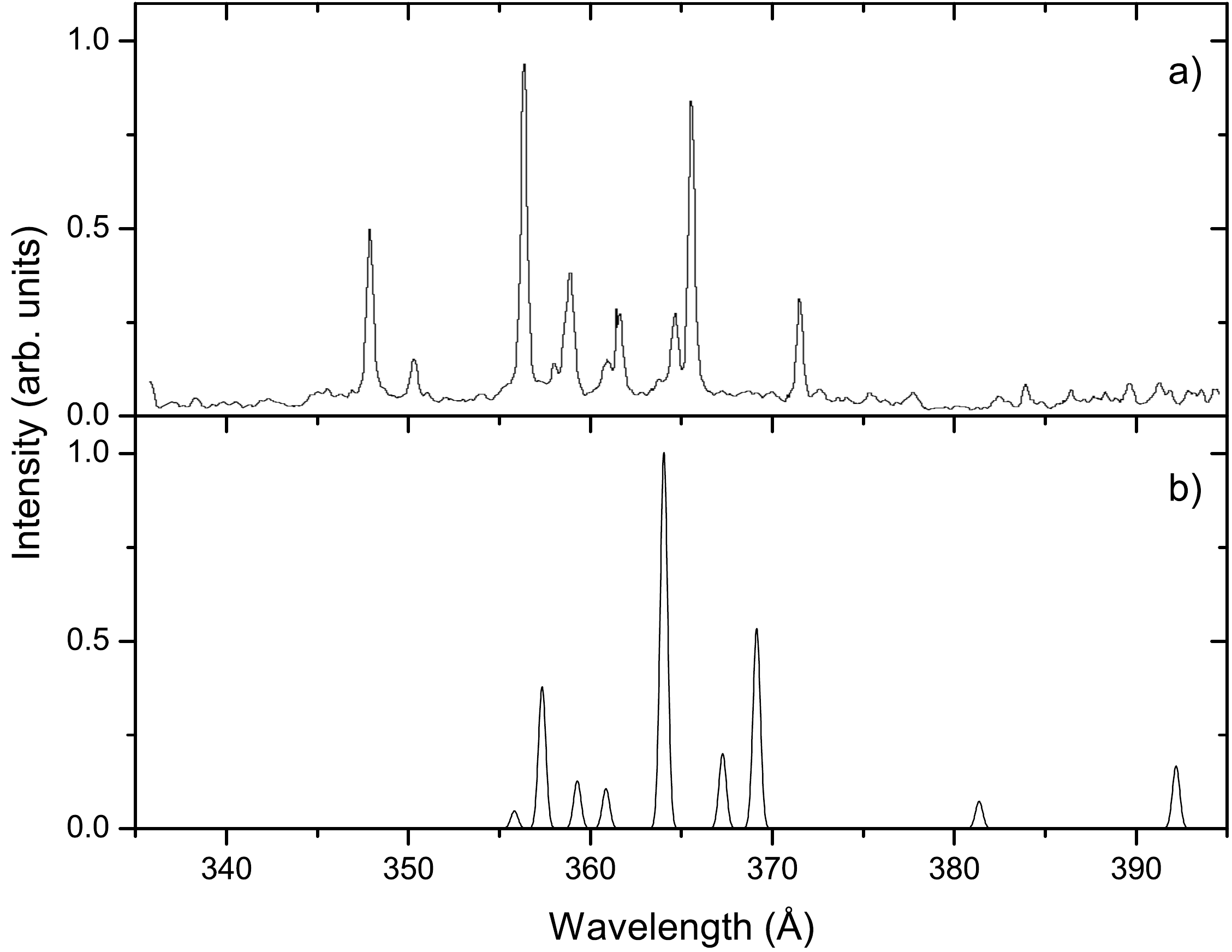}%
 \caption{\label{f1} Comparison of a) experimental \cite{Hutton_2003nimb_205_114} and b) theoretical spectra. { Lines in theoretical spectrum originate from the $4f^{13} 5s 5p \rightarrow  4f^{13} 5s^{2}$ transition.  Theoretical spectrum corresponds to coronal approximation ($E_{e}=300$ eV, $n_e=10^{12}$ cm$^{-3}$) when population of levels considered from the ground level of $W^{13+}$.} }
 \end{figure}

Theoretical spectra which correspond to the radiative transitions after population of levels by electron impact excitation from the ground level {( Eq.(\ref{e1}) with Eq. (\ref{e2}))} at various beam energies contain two prominent groups of lines { (Fig. \ref{f01})}.  One group is obtained at 220 -- 240 \AA{} while the other is formed at  350 -- 370 \AA{}. { The theoretical spectra are convoluted with a Gaussian function with a half width of 0.2 \AA{} to simulate experimental resolution.} In this case, further radiative decay of the levels through the cascade emission  was not taken into account while obtaining those theoretical spectra. The structure of lines in the spectrum is similar to the one observed in the EBIT experiment \cite{Hutton_2003nimb_205_114, 2008_Wu_cjp_86_125}. 
{ The wavelength of the strongest line of the second group is equal to 364.04 \AA{} which very well agrees with the experimental wavelength of 365.3 \AA{} (Fig. \ref{f1}).}
However,  in the previous work \cite{Hutton_2003nimb_205_114} the strongest line of the experimental spectrum  was identified as $5p \rightarrow 5s$ transition while our data show that the line corresponds to $4f^{13} 5s 5p \rightarrow 4f^{13} 5s^{2}$ transition. We note that $4f^{13} 5s 5p$ configuration is the first excited configuration connected to the ground one through strong electric dipole transitions.  

{ Unfortunately, the wavelengths of the first group of lines (220 -- 240 \AA{}) significantly differ from the experimental values. 
Our CI basis consists of one- and/or two-electron excitations from $4f^{13} 5s^{2}$, $4f^{14} 5s$ and $4f^{13} 5s 5p$ configurations.  Thus, good agreement for theoretical wavelengths of the second group of lines (350 -- 370 \AA{}) with observed data indicates that main correlation effects important for $4f^{13} 5s 5p \rightarrow  4f^{13} 5s^{2}$ transition are taken into account. Therefore, we suggest that wavelengths at 250 -- 270 \AA{} in the experimental spectrum are formed by other transitions in $W^{13+}$. 
Consequently, we have analyzed population of the lowest metastable levels of $4f^{12} 5s^{2} 5p$ configuration. }

 \begin{table}
 \caption{\label{t1} Energies $E$ (in eV) and radiative lifetimes $\tau$ (in s) of the lowest excited levels of $W^{13+}$. 
Energies are given relative to the energy of the ground level. Closed inner subshells are omitted in the notations of the levels.}
 \begin{tabular}{llrll}
\hline
 No. & $E$, eV & $J$  & Level & $\tau$, s \\
 \hline
 0  & 0.0    & 7/2  & $4f_{7/2}^7$(7/2) $5s^{2}$ &   -- \\
 1  & 2.1848 & 5/2  & $4f_{5/2}^5$(5/2) $5s^{2}$ &   $1.19\times10^{-2}$ \\
 2  & 3.8279 & 11/2 & $4f_{7/2}^6$(6) $5s^{2}$ $5p_{1/2}$ &  $3.30\times10^{2}$ \\
 3  & 4.4857 & 13/2 & $4f_{7/2}^6$(6) $5s^{2}$ $5p_{1/2}$ &  $2.64\times10^{0}$ \\
 4  & 5.1054 & 7/2  & $4f_{7/2}^6$(4) $5s^{2}$ $5p_{1/2}$   &  $2.75\times10^{0}$ \\
 5  & 5.4559 & 9/2  & $4f_{7/2}^6$(4) $5s^{2}$ $5p_{1/2}$    &  $3.55\times10^{-1}$ \\
 6  & 6.2594 & 11/2 & $4f_{5/2}^5$(5/2)  $4f_{7/2}^7$(7/2)5 $5s^{2}$ $5p_{1/2}$ &  $1.16\times10^{-2}$ \\
 7  & 6.3337 & 9/2  & $4f_{5/2}^5$(5/2)  $4f_{7/2}^7$(7/2)5 $5s^{2}$ $5p_{1/2}$  &  $5.40\times10^{-3}$ \\
 8  & 7.0263 & 7/2  & $4f_{5/2}^5$(5/2)  $4f_{7/2}^7$(7/2)4 $5s^{2}$ $5p_{1/2}$   &  $2.36\times10^{-2}$ \\
 9  & 7.3568 & 5/2  & $4f_{5/2}^5$(5/2)  $4f_{7/2}^7$(7/2)3 $5s^{2}$ $5p_{1/2}$   &  $2.96\times10^{-2}$ \\ 
 10 & 7.4865 & 9/2  & $4f_{5/2}^5$(5/2)  $4f_{7/2}^7$(7/2)4 $5s^{2}$ $5p_{1/2}$   &  $5.67\times10^{-3}$ \\
 11 & 7.5391 & 3/2  & $4f_{7/2}^6$(2) $5s^{2}$  $5p_{1/2}$                        &    $2.15\times10^{0}$ \\
 12 & 7.8546 & 5/2  & $4f_{5/2}^5$(5/2)  $4f_{7/2}^7$(7/2)3 $5s^{2}$ $5p_{1/2}$   &  $7.49\times10^{-3}$ \\
 13 & 8.2631 & 7/2  & $4f_{5/2}^5$(5/2)  $4f_{7/2}^7$(7/2)3 $5s^{2}$ $5p_{1/2}$   &  $4.97\times10^{-3}$ \\
 \hline
 \end{tabular}

 \end{table}

The lowest level of $4f^{12} 5s^{2} 5p$ configuration { compared with the ground one} gains much larger relative population after excitation from the ground level when Eq. (\ref{e1}) with Eq. (\ref{e2}) are used to find population of excited levels in corona model. Collisional-radiative modeling ($n_e=10^{12}$ cm$^{-3}$, $E_e=200$ eV) performed for the twelve lowest levels of $W^{13+}$ ion { (Table \ref{t1})} shows that the first level of $4f^{12} 5s^{2} 5p$ configuration has also the largest population.   As well, the levels excited from the ground level  decay through the cascade emission mainly to the lowest level of $4f^{12} 5s^{2} 5p$ configuration if weak transitions from the levels of $4f^{12} 5s^{2} 5p$ configuration to the lower levels of the ground configuration are not taken into account. On the other hand, the cascade emission after the excitation from the lowest level of $4f^{12} 5s^{2} 5p$ configuration populates mainly the second level of this configuration.  

Table \ref{t1} presents energy levels and radiative lifetimes for 14 lowest levels of $W^{13+}$ ion. The levels of $4f^{12} 5s^{2} 5p$ configuration have the largest lifetimes { among the excited levels of the five lowest configurations for which extended study of radiative transitions (E1, E2, E3, M1, M2) are performed}. The lowest level of the configuration can decay to the ground level through electric quadrupole transition. The large lifetimes of these levels and population of the levels through the cascade emission { indicate that population of the lowest levels of $4f^{12} 5s^{2} 5p$ configuration exceeds population of the ground configuration.} Thus, { mainly}  excitations from the metastable levels of the first excited configuration are responsible for line formation in the EBIT spectra of $W^{13+}$ ion.   These  levels have to be considered in the coronal approximation. 

\begin{figure}
 \includegraphics[scale=0.3]{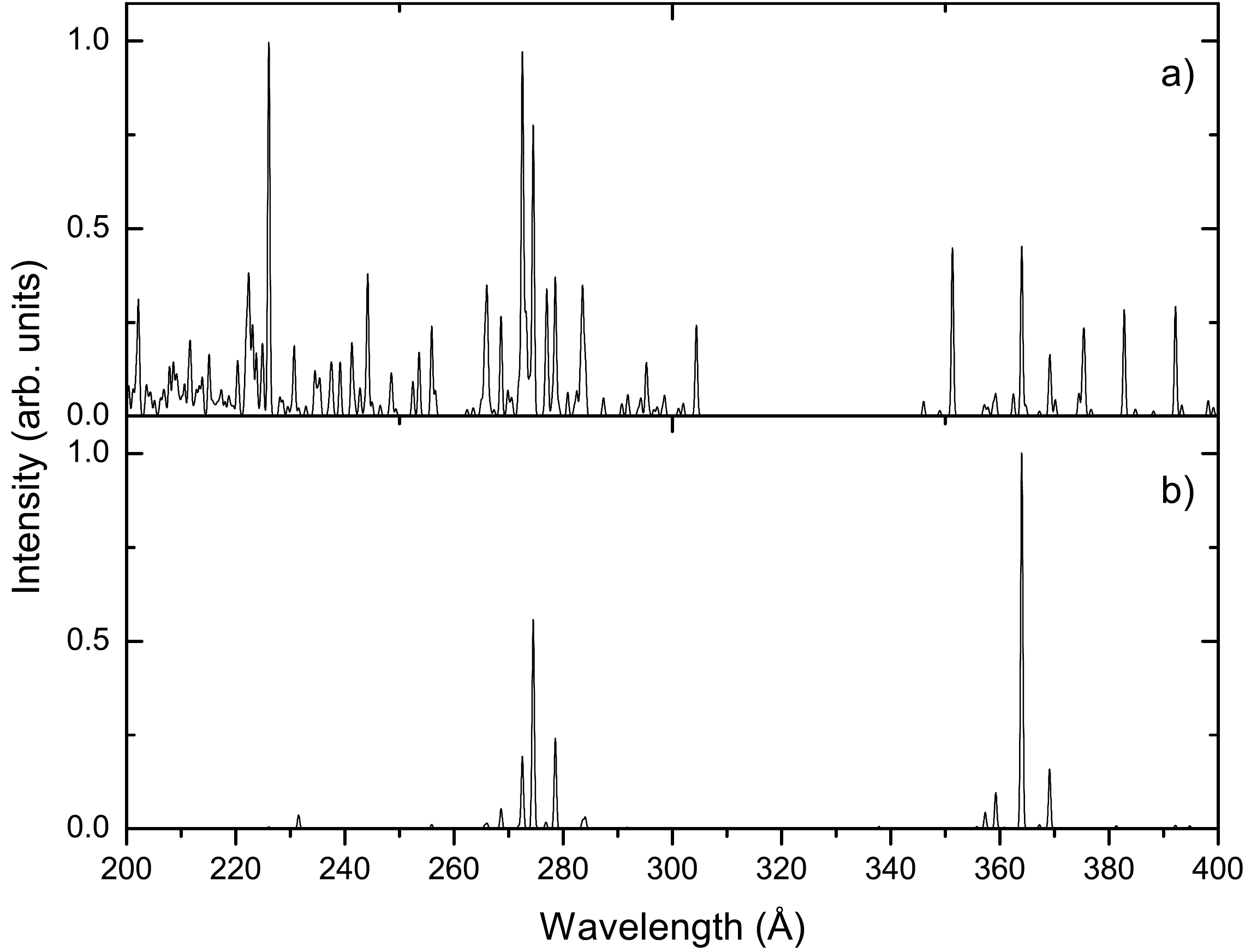}%
 \caption{\label{f2} 
Spectra obtained a) in the coronal approximation after excitation from the $4f_{7/2}^6(6) 5s^{2} 5p_{1/2} J=13/2$ level ($E_{e}=200$ eV, $n_e=10^{12}$ cm$^{-3}$) and b) further decay through the cascade emission. }
 \end{figure}

Fig. \ref{f2}a displays the theoretical spectrum obtained after electron impact excitation from the  second lowest level of $4f^{12} 5s^{2} 5p$ configuration at electron density equal to $10^{12}$ cm$^{-3}$ and beam energy of $200$ eV . 
Only the cascade emission after the excitation from the level again reveals  two groups of lines (Fig. \ref{f2}b). { The similar situation is observed with excitation from the lowest level of $4f^{12} 5s^{2} 5p$ configuration.}

One group of lines is formed at 360--370 \AA{} as in the case of excitation from the ground level. It is interesting that those lines again correspond to the transitions from the levels of $4f^{13} 5s 5p$ configuration to the ground configuration. The initial level of the transition which corresponds to the strongest line at 364.04 \AA{} is populated by the weak electric dipole transitions from the levels of $4f^{11} 5s^{2} 5p^{2}$ configuration during the cascade emission. Those transitions are available due to configuration mixing and therefore transition wavelengths and probabilities are very sensitive to the used set of interacting configurations.

The other group of lines appears at 260--280 \AA{}. These lines correspond to  $4f^{12} 5s 5p^{2} \rightarrow 4f^{12} 5s^{2} 5p$ transition. Consequently, lines of the experimental spectra  can be identified as electric dipole  transition of $5p \rightarrow 5s $ type. However, wavelengths of those lines differs by about 16.5 \AA{} from the experimental group of lines obtained at 258 \AA{}. 
Correlation effects for the levels of $4f^{12} 5s^{2} 5p$ and $4f^{12} 5s 5p^{2}$ configurations are very important and it requires large computer resources for such large number of levels which have to be included in CI basis. 
{ Large discrepancies for wavelengths of this group of lines can be explained by the fact that in CI basis we included only one- and/or two-electron excitations for $4f^{13} 5s^{2}$, $4f^{14} 5s$ and $4f^{13} 5s 5p$ configurations but not for $4f^{12} 5s^{2} 5p$ and $4f^{12} 5s 5p^{2}$ configurations.} On the other hand, for Er-like tungsten it was found that FAC shows discrepancies within 20 \AA{} of the measured values due to large configuration-interaction effects \cite{2010jpb_43_144009_clementson}.

Levels of $4f^{12} 5s 5p^{2}$  configuration are mainly populated through cascade emission from levels of $4f^{11} 5s^{2} 5p^{2}$ configuration. Those transitions are only available due to configuration interaction effects and are very weak. They correspond to electric octupole transitions in the single configuration approach.

\begin{figure}
 \includegraphics[scale=0.3]{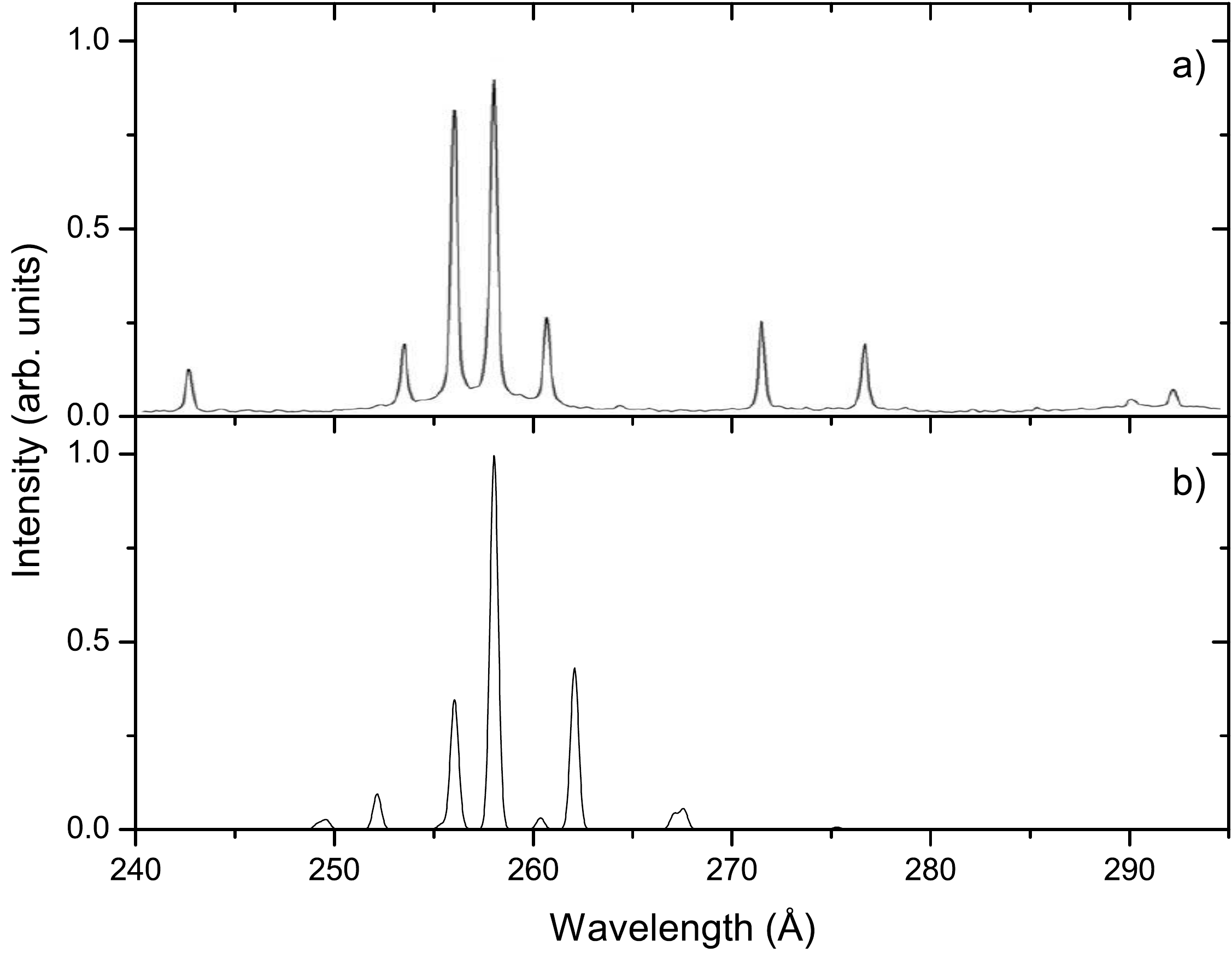}%
 \caption{\label{f3} 
Comparison of a) experimental \cite{2008_Wu_cjp_86_125} and b) theoretical spectra. Lines in theoretical spectrum originate from the $4f^{12} 5s 5p^{2} \rightarrow  4f^{12} 5s^{2} 5p$ transition. Theoretical spectrum was shifted to the short wavelength side by 16.5 \AA{} for better agreement with the experimental values. { Theoretical spectrum corresponds to emission cascade after coronal approximation ($E_{e}=200$ eV, $n_e=10^{12}$ cm$^{-3}$) applied to find initial population of levels excited from the $4f_{7/2}^6(6) 5s^{2} 5p_{1/2} J=13/2$ level of $W^{13+}$. }                                                                                                                                                                                                                                                                                                                            }
 \end{figure}

Correspondence among theoretical and experimental lines for the group of lines is presented in Fig. \ref{f3}. Theoretical spectrum in Fig. \ref{f3}b was shifted to the short wavelength side by 16.5 \AA{} for a better agreement with experimental lines.  

Finaly, we checked what gives charge exchange (CX) of the ions with neutral atoms that are present in the EBIT to the formation of spectrum for $W^{13+}$ ion. Due to CX of neutrals with highly charged ions, electrons are mainly captured to the excited levels with principal quantum number $n \approx Z^{0.75}_{eff}$ ($Z_{eff}$ is effective charge of the ion) \cite{janev1985}. For $W^{13+}$, it gives $n=6.8$. We calculated spectrum which corresponds to the cascade emission from levels with $n=6$ taking population of levels equal to their statistical weights. Complex structure of lines in the theoretical spectrum shows that influence of CX to the line intensities of experimental spectrum is small. Furthermore, good agreement obtained for relative line intesities when CX process was not included in CRM for higher ionization charges \cite{2006pra_74_042514_Ralchenko, 2007jpb_40_3861_ralchenko} { demonstrates that charge exchange is not important for line formation in EBIT plasma}.

\section{Conclusions}

We showed that cascade emission has to be taken into account in combination with coronal approximation for the analysis of spectral lines from the electron beam ion trap plasma in electron trapping mode. Our demonstration is based on the investigation of line intensities obtained after excitation from the ground or metastable levels with consequent cascade emission for $W^{13+}$ ion. Cascade emission heightens only a few lines from the complex structure of lines which obtained using corona model. 
For the first time, lines in the experimental spectrum are identified as $4f^{13} 5s 5p \rightarrow 4f^{13} 5s^{2}$ and $4f^{12} 5s 5p^{2} \rightarrow 4f^{12} 5s^{2} 5p$ transitions in $W^{13+}$ ion. Good agreement is obtained between experimental and theoretical spectra only after cascade emission was included in calculations. 

Further investigation is needed for wavelengths and relative line intensities in order to obtain better agreement to the experimental spectrum. 
Large scale calculations in Dirac-Fock \cite{grant2007relativistic} approach  with selected CI basis  from configuration interaction strength \cite{2010pra_81_012506_jonauskas} are on the way for $4f^{13} 5s 5p \rightarrow 4f^{13} 5s^{2}$ transition and will be published elsewhere.

\section*{Akcnowledgement}
This research was funded by European Social Fund under the Global Grant Measure (No.: VP1-3.1-\v{S}MM-07-K-02-015).



\bibliographystyle{elsarticle-num-names}
\bibliographystyle{model3a-num-names}

\begin{thebibliography}{27}
\providecommand{\natexlab}[1]{#1}
\providecommand{\url}[1]{\texttt{#1}}
\providecommand{\urlprefix}{URL }
\expandafter\ifx\csname urlstyle\endcsname\relax
  \providecommand{\doi}[1]{doi:\discretionary{}{}{}#1}\else
  \providecommand{\doi}{doi:\discretionary{}{}{}\begingroup
  \urlstyle{rm}\Url}\fi
\providecommand{\eprint}[2][]{\url{#2}}
\providecommand{\bibinfo}[2]{#2}
\ifx\xfnm\undefined \def\xfnm[#1]{\unskip,\space#1}\fi
\makeatletter\def\@biblabel#1{#1.}\makeatother

\bibitem[{Marrs et~al.(1988)Marrs, Levine, Knapp and
  Henderson}]{1988prl_60_1715_marrs}
\bibinfo{author}{Marrs\xfnm[ R.E.]}, \bibinfo{author}{Levine\xfnm[ M.A.]},
  \bibinfo{author}{Knapp\xfnm[ D.A.]}, \bibinfo{author}{Henderson\xfnm[ J.R.]}.
\newblock \bibinfo{title}{Measurement of electron-impact--excitation cross sections for very highly charged ions}.
\newblock \emph{\bibinfo{journal}{Phys Rev Lett}}
  \bibinfo{year}{1988};\hspace{0pt}\textbf{\bibinfo{volume}{60}}:\bibinfo{pages}{1715--18}.

\bibitem[{Ralchenko et~al.({2006})Ralchenko, Tan, Gillaspy, Pomeroy and
  Silver}]{2006pra_74_042514_Ralchenko}
\bibinfo{author}{Ralchenko\xfnm[ Y.]}, \bibinfo{author}{Tan\xfnm[ J.N.]},
  \bibinfo{author}{Gillaspy\xfnm[ J.D.]}, \bibinfo{author}{Pomeroy\xfnm[ J.M.]}, \bibinfo{author}{Silver\xfnm[ E.]}.
\newblock \bibinfo{title}{{Accurate modeling of benchmark X-ray spectra from highly charged ions of tungsten}}.
\newblock \emph{\bibinfo{journal}{Phys Rev A}}
  \bibinfo{year}{{2006}};\hspace{0pt}\textbf{\bibinfo{volume}{{74}}}(\bibinfo{number}{{4}}):\bibinfo{pages}{042514}.

\bibitem[{Ralchenko et~al.(2007)Ralchenko, Reader, Pomeroy, Tan and
  Gillaspy}]{2007jpb_40_3861_ralchenko}
\bibinfo{author}{Ralchenko\xfnm[ Y.]}, \bibinfo{author}{Reader\xfnm[ J.]},
  \bibinfo{author}{Pomeroy\xfnm[ J.M.]}, \bibinfo{author}{Tan\xfnm[ J.N.]},
  \bibinfo{author}{Gillaspy\xfnm[ J.D.]}.
\newblock \bibinfo{title}{Spectra of $W^{39+}$-$W^{47+}$ in the 12-20 nm region observed
  with an EBIT light source}.
\newblock \emph{\bibinfo{journal}{J Phys B}}
  \bibinfo{year}{2007};\hspace{0pt}\textbf{\bibinfo{volume}{40}}(\bibinfo{number}{19}):\bibinfo{pages}{3861--75}.

\bibitem[{Radtke et~al.(2007)Radtke, Biedermann, Fussmann, Schwob, Mandelbaum
  and Doron}]{2007apmidf_13_45_radtke}
\bibinfo{author}{Radtke\xfnm[ R.]}, \bibinfo{author}{Biedermann\xfnm[ C.]},
  \bibinfo{author}{Fussmann\xfnm[ G.]}, \bibinfo{author}{Schwob\xfnm[ J.]},
  \bibinfo{author}{Mandelbaum\xfnm[ P.]}, \bibinfo{author}{Doron\xfnm[ R.]}.
\newblock \bibinfo{title}{Measured line spectra and calculated atomic physics
  data for highly charged tungsten ions}.
\newblock \emph{\bibinfo{journal}{Atomic and Plasma - Material Interaction Data
  for Fusion}}
  \bibinfo{year}{2007};\hspace{0pt}\textbf{\bibinfo{volume}{13}}:\bibinfo{pages}{45--66}.

\bibitem[{Ralchenko et~al.(2008)Ralchenko, Draganic, Tan, Gillaspy, Pomeroy,
  Reader et~al.}]{2008jpb_41_021003_ralchenko}
\bibinfo{author}{Ralchenko\xfnm[ Y.]}, \bibinfo{author}{Draganic\xfnm[ I.N.]},
  \bibinfo{author}{Tan\xfnm[ J.N.]}, \bibinfo{author}{Gillaspy\xfnm[ J.D.]},
  \bibinfo{author}{Pomeroy\xfnm[ J.M.]}, \bibinfo{author}{Reader\xfnm[ J.]},
  et~al.
\newblock \bibinfo{title}{EUV spectra of highly-charged ions $W^{54+}$-$W^{63+}$ relevant to ITER diagnostics}.
\newblock \emph{\bibinfo{journal}{J Phys B}}
  \bibinfo{year}{2008};\hspace{0pt}\textbf{\bibinfo{volume}{41}}(\bibinfo{number}{2}):\bibinfo{pages}{021003
  (6pp)}.


\bibitem[{Chen({2008})}]{2008mnras_386_l62_chen}
\bibinfo{author}{Chen\xfnm[ G.X.]}.
\newblock \bibinfo{title}{{X-ray line ratio 3C/3D in Fe XVII}}.
\newblock \emph{\bibinfo{journal}{{Mon Not R Astron Soc}}}
  \bibinfo{year}{{2008}};\hspace{0pt}\textbf{\bibinfo{volume}{{386}}}(\bibinfo{number}{{1}}):\bibinfo{pages}{L62--L66}.

\bibitem[{Jonauskas et~al.(2007)Jonauskas, Ku\v{c}as and
  Karazija}]{Jonauskas2007jpb_40_2179}
\bibinfo{author}{Jonauskas\xfnm[ V.]}, \bibinfo{author}{Ku\v{c}as\xfnm[ S.]},
  \bibinfo{author}{Karazija\xfnm[ R.]}.
\newblock \bibinfo{title}{On the interpretation of the intense emission of
  tungsten ions at about 5 nm}.
\newblock \emph{\bibinfo{journal}{J Phys B}}
  \bibinfo{year}{2007};\hspace{0pt}\textbf{\bibinfo{volume}{40}}(\bibinfo{number}{11}):\bibinfo{pages}{2179--88}.


\bibitem[{Jonauskas et~al.(2003)Jonauskas, Partanen, Ku\v{c}as, Karazija,
  Huttula, Aksela et~al.}]{2003jpb_36_4403_jonauskas}
\bibinfo{author}{Jonauskas\xfnm[ V.]}, \bibinfo{author}{Partanen\xfnm[ L.]},
  \bibinfo{author}{Ku\v{c}as\xfnm[ S.]}, \bibinfo{author}{Karazija\xfnm[ R.]},
  \bibinfo{author}{Huttula\xfnm[ M.]}, \bibinfo{author}{Aksela\xfnm[ S.]},
  et~al.
\newblock \bibinfo{title}{Auger cascade satellites following 3d ionization in
  xenon}.
\newblock \emph{\bibinfo{journal}{J Phys B}}
  \bibinfo{year}{2003};\hspace{0pt}\textbf{\bibinfo{volume}{36}}(\bibinfo{number}{22}):\bibinfo{pages}{4403--16}.


\bibitem[{Jonauskas et~al.(2008)Jonauskas, Karazija and
  Ku\v{c}as}]{2008jpb_41_215005_jonauskas}
\bibinfo{author}{Jonauskas\xfnm[ V.]}, \bibinfo{author}{Karazija\xfnm[ R.]},
  \bibinfo{author}{Ku\v{c}as\xfnm[ S.]}.
\newblock \bibinfo{title}{The essential role of many-electron auger transitions
  in the cascades following the photoionization of 3p and 3d shells of kr}.
\newblock \emph{\bibinfo{journal}{J Phys B}}
  \bibinfo{year}{2008};\hspace{0pt}\textbf{\bibinfo{volume}{41}}(\bibinfo{number}{21}):\bibinfo{pages}{215005
  (5pp)}.



\bibitem[{Palaudoux et~al.({2010})Palaudoux, Lablanquie, Andric, Ito,
  Shigemasa, Eland et~al.}]{2010pra_82_043419_palaudoux}
\bibinfo{author}{Palaudoux\xfnm[ J.]}, \bibinfo{author}{Lablanquie\xfnm[ P.]},
  \bibinfo{author}{Andric\xfnm[ L.]}, \bibinfo{author}{Ito\xfnm[ K.]},
  \bibinfo{author}{Shigemasa\xfnm[ E.]}, \bibinfo{author}{Eland\xfnm[ J.H.D.]},
  et~al.
\newblock \bibinfo{title}{{Multielectron spectroscopy: Auger decays of the
  krypton 3d hole}}.
\newblock \emph{\bibinfo{journal}{{Phys Rev A}}}
  \bibinfo{year}{{2010}};\hspace{0pt}\textbf{\bibinfo{volume}{{82}}}(\bibinfo{number}{{4}}):\bibinfo{pages}{043419}.

\bibitem[{Jonauskas et~al.(2011)Jonauskas, Ku\ifmmode~\check{c}\else
  \v{c}\fi{}as and Karazija}]{2011pra_84_053415_jonauskas}
\bibinfo{author}{Jonauskas\xfnm[ V.]},
  \bibinfo{author}{Ku\ifmmode~\check{c}\else \v{c}\fi{}as\xfnm[ S.]},
  \bibinfo{author}{Karazija\xfnm[ R.]}.
\newblock \bibinfo{title}{Auger decay of 3$p$-ionized krypton}.
\newblock \emph{\bibinfo{journal}{Phys Rev A}}
  \bibinfo{year}{2011};\hspace{0pt}\textbf{\bibinfo{volume}{84}}:\bibinfo{pages}{053415}.


\bibitem[{Currell(2001)}]{currel2001}
\bibinfo{author}{Currell\xfnm[ F.J.]}.
\newblock \emph{\bibinfo{title}{Trapping Highly Charged Ions: Fundamentals and
  Applications}}.
\newblock \bibinfo{publisher}{Nova Science Publishers, New York};
  \bibinfo{year}{2001}, \hspace{0pt}p.~\bibinfo{pages}{3}.

\bibitem[{Gillaspy et~al.({1995})Gillaspy, Aglitskiy, Bell, Brown, Chantler,
  Deslattes et~al.}]{1995ps_59_392_gillaspy}
\bibinfo{author}{Gillaspy\xfnm[ J.]}, \bibinfo{author}{Aglitskiy\xfnm[ Y.]},
  \bibinfo{author}{Bell\xfnm[ E.]}, \bibinfo{author}{Brown\xfnm[ C.]},
  \bibinfo{author}{Chantler\xfnm[ C.]}, \bibinfo{author}{Deslattes\xfnm[ R.]},
  et~al.
\newblock \bibinfo{title}{{Overview of the electron-beam ion-trap program at NIST}}.
\newblock \emph{\bibinfo{journal}{{Physica Scripta}}}
  \bibinfo{year}{{1995}};\hspace{0pt}\textbf{\bibinfo{volume}{{T59}}}:\bibinfo{pages}{392--5}.


\bibitem[{Liang et~al.({2009})Liang, Lopez-Urrutia, Baumann, Epp, Gonchar,
  Lapierre et~al.}]{2009apj_702_838_liang}
\bibinfo{author}{Liang\xfnm[ G.Y.]}, \bibinfo{author}{Lopez-Urrutia\xfnm[
  J.R.C.]}, \bibinfo{author}{Baumann\xfnm[ T.M.]}, \bibinfo{author}{Epp\xfnm[
  S.W.]}, \bibinfo{author}{Gonchar\xfnm[ A.]}, \bibinfo{author}{Lapierre\xfnm[
  A.]}, et~al.
\newblock \bibinfo{title}{{
Experimental investigations of ion charge
  distributions, effective electron densities, and electron-ion cloud overlap
  in electron beam ion trap plasma using extreme-ultraviolet spectroscopy}}.
\newblock \emph{\bibinfo{journal}{{Astrophys J}}}
  \bibinfo{year}{{2009}};\hspace{0pt}\textbf{\bibinfo{volume}{{702}}}(\bibinfo{number}{{2}}):\bibinfo{pages}{838--50}.

\bibitem[{Chen et~al.(2004)Chen, Beiersdorfer, Heeter, Liedahl, Naranjo-Rivera,
  Tr\"{a}bert, Gu, and Lepson}]{2004apj_611_598_chen}
\bibinfo{author}{H.~Chen}, \bibinfo{author}{P.~Beiersdorfer},
  \bibinfo{author}{L.~A. Heeter}, \bibinfo{author}{D.~A. Liedahl},
  \bibinfo{author}{K.~L. Naranjo-Rivera}, \bibinfo{author}{E.~TrÃÂ¤bert},
  \bibinfo{author}{M.~F. Gu}, \bibinfo{author}{J.~K. Lepson},
  \bibinfo{title}{Experimental and Theoretical Evaluation of Density-sensitive
  N VI, Ar XIV, and Fe XXII Line Ratios}, 
\newblock \emph{\bibinfo{journal}{{Astrophys J}}}
  \bibinfo{year}{{2004}};\hspace{0pt}\textbf{\bibinfo{volume}{{611}}}(\bibinfo{number}{{1}}):\bibinfo{pages}{598--604}.



\bibitem[{Beiersdorfer et~al.(1996)Beiersdorfer, Schweikhard, López-Urrutia
  and Widmann}]{1996rsi_67_11_Beiersdorfer}
\bibinfo{author}{Beiersdorfer\xfnm[ P.]}, \bibinfo{author}{Schweikhard\xfnm[
  L.]}, \bibinfo{author}{López-Urrutia\xfnm[ J.C.]},
  \bibinfo{author}{Widmann\xfnm[ K.]}.
\newblock \bibinfo{title}{{The magnetic trapping mode of an electron beam ion
  trap: New opportunities for highly charged ion research}}.
\newblock \emph{\bibinfo{journal}{RSI}}
  \bibinfo{year}{1996};\hspace{0pt}\textbf{\bibinfo{volume}{67}}(\bibinfo{number}{11}):\bibinfo{pages}{3818}.

\bibitem[{{Hutton} et~al.(2003){Hutton}, {Zou}, {Reyna Almandos}, {Biedermann},
  {Radtke}, {Greier} et~al.}]{Hutton_2003nimb_205_114}
\bibinfo{author}{{Hutton}\xfnm[ R.]}, \bibinfo{author}{{Zou}\xfnm[ Y.]},
  \bibinfo{author}{{Reyna Almandos}\xfnm[ J.]},
  \bibinfo{author}{{Biedermann}\xfnm[ C.]}, \bibinfo{author}{{Radtke}\xfnm[
  R.]}, \bibinfo{author}{{Greier}\xfnm[ A.]}, et~al.
\newblock \bibinfo{title}{EBIT spectroscopy of Pm-like tungsten}.
\newblock \emph{\bibinfo{journal}{Nucl Instr Meth B}}
  \bibinfo{year}{2003};\hspace{0pt}\textbf{\bibinfo{volume}{205}}:\bibinfo{pages}{114--18}.

\bibitem[{Wu and Hutton({2008})}]{2008_Wu_cjp_86_125}
\bibinfo{author}{Wu\xfnm[ S.]}, \bibinfo{author}{Hutton\xfnm[ R.]}.
\newblock \bibinfo{title}{{Applications of EBITs to spectra of multi-electron
  ions: some solved and some unsolved problems}}.
\newblock \emph{\bibinfo{journal}{{Can J Phys}}}
  \bibinfo{year}{{2008}};\hspace{0pt}\textbf{\bibinfo{volume}{{86}}}:\bibinfo{pages}{125--29}.

\bibitem[{{Cowan}(1981)}]{Cowan_1981tass.book.....C}
\bibinfo{author}{{Cowan}\xfnm[ R.D.]}.
\newblock \emph{\bibinfo{title}{{The Theory of Atomic Structure and Spectra}}}.
\newblock \bibinfo{publisher}{Berkeley, CA: University of California Press};
  \bibinfo{year}{1981}.

\bibitem[{{Tr{\"a}bert} et~al.({2009}){Tr{\"a}bert}, Vilkas and
  Ishikawa}]{Trabert_2008jpcs_163_012017}
\bibinfo{author}{{Tr{\"a}bert}\xfnm[ E.]}, \bibinfo{author}{Vilkas\xfnm[
  M.J.]}, \bibinfo{author}{Ishikawa\xfnm[ Y.]}.
\newblock \bibinfo{title}{{A tale of two lines: Searching for the 5s-5p
  resonance lines in Pm-like ion spectra}}.
\newblock \emph{\bibinfo{journal}{{J Phys Conf Ser}}}
  \bibinfo{year}{{2009}};\hspace{0pt}\textbf{\bibinfo{volume}{{163}}}:\bibinfo{pages}{012017}.

\bibitem[{{{Curtis}, L.~J.} and {{Ellis},
  D.~G.}(1980)}]{Curtis_1980prl_45_2099}
\bibinfo{author}{{{Curtis}, L.~J.}\xfnm[]}, \bibinfo{author}{{{Ellis},
  D.~G.}\xfnm[]}.
\newblock \bibinfo{title}{Alkalilike spectra in the promethium isoelectronic sequence}.
\newblock \emph{\bibinfo{journal}{Phys Rev Lett}}
  \bibinfo{year}{1980};\hspace{0pt}\textbf{\bibinfo{volume}{45}}:\bibinfo{pages}{2099--102}.


\bibitem[{Vilkas et~al.(2008)Vilkas, Ishikawa and
  Tr{\"a}bert}]{Vilkas_2008pra_77_042510}
\bibinfo{author}{Vilkas\xfnm[ M.J.]}, \bibinfo{author}{Ishikawa\xfnm[ Y.]},
  \bibinfo{author}{Tr{\"a}bert\xfnm[ E.]}.
\newblock \bibinfo{title}{Electric-dipole $5s-5p$ transitions in promethiumlike
  ions}.
\newblock \emph{\bibinfo{journal}{Phys Rev A}}
  \bibinfo{year}{2008};\hspace{0pt}\textbf{\bibinfo{volume}{77}}:\bibinfo{pages}{042510}.

\bibitem[{{Gu}(2003)}]{Gu_2003ApJ...582.1241G}
\bibinfo{author}{{Gu}\xfnm[ M.F.]}.
\newblock \bibinfo{title}{{Indirect X-Ray Line-Formation Processes in Iron
  L-Shell Ions}}.
\newblock \emph{\bibinfo{journal}{Astrophys J}}
  \bibinfo{year}{2003};\hspace{0pt}\textbf{\bibinfo{volume}{582}}:\bibinfo{pages}{1241--50}.


\bibitem[{Clementson et~al.(2010)Clementson, Beiersdorfer, Magee, McLean and
  Wood}]{2010jpb_43_144009_clementson}
\bibinfo{author}{Clementson\xfnm[ J.]}, \bibinfo{author}{Beiersdorfer\xfnm[
  P.]}, \bibinfo{author}{Magee\xfnm[ E.W.]}, \bibinfo{author}{McLean\xfnm[
  H.S.]}, \bibinfo{author}{Wood\xfnm[ R.D.]}.
\newblock \bibinfo{title}{{Tungsten spectroscopy relevant to the diagnostics of
  ITER divertor plasmas}}.
\newblock \emph{\bibinfo{journal}{Journal of Physics B: Atomic, Molecular and
  Optical Physics}}
  \bibinfo{year}{2010};\hspace{0pt}\textbf{\bibinfo{volume}{43}}(\bibinfo{number}{14}):\bibinfo{pages}{144009}.

\bibitem[{Janev et~al.(1985)Janev, Presnyakov and Shevelko}]{janev1985}
\bibinfo{author}{Janev\xfnm[ R.K.]}, \bibinfo{author}{Presnyakov\xfnm[ L.P.]},
  \bibinfo{author}{Shevelko\xfnm[ V.P.]}.
\newblock \emph{\bibinfo{title}{Physics of Highly Charged Ions}}.
\newblock \bibinfo{publisher}{Berlin: Springer}; \bibinfo{year}{1985}.

\bibitem[{Grant(2007)}]{grant2007relativistic}
\bibinfo{author}{Grant\xfnm[ I.]}.
\newblock \emph{\bibinfo{title}{Relativistic quantum theory of atoms and
  molecules: theory and computation}}.
\newblock Springer series on atomic, optical, and plasma physics.
  \bibinfo{publisher}{Springer}; \bibinfo{year}{2007}.

\bibitem[{Jonauskas et~al.(2010)Jonauskas, Kisielius, {Kynien\. e}, {S.~Ku{\v
  {c}}as} and Norrington}]{2010pra_81_012506_jonauskas}
\bibinfo{author}{Jonauskas\xfnm[ V.]}, \bibinfo{author}{Kisielius\xfnm[ R.]},
  \bibinfo{author}{{Kynien\. e}\xfnm[ A.]}, \bibinfo{author}{{S.~Ku{\v
  {c}}as}\xfnm[]}, \bibinfo{author}{Norrington\xfnm[ P.H.]}.
\newblock \bibinfo{title}{Magnetic dipole transitions in 4d$^n$ configurations
  of tungsten ions}.
\newblock \emph{\bibinfo{journal}{Phys Rev A}}
  \bibinfo{year}{2010};\hspace{0pt}\textbf{\bibinfo{volume}{81}}:\bibinfo{pages}{012506}.

\end{thebibliography}











\end{document}